\documentclass[aps,notitlepage,a4paper,superscriptaddress,11pt]{article}

\usepackage[utf8]{inputenc}
\usepackage[dvips]{color}
\usepackage[toc]{appendix}
\usepackage[hang,flushmargin]{footmisc} 
\usepackage{titlesec,titletoc,authblk,abstract,latexsym,graphicx,microtype,booktabs,cleveref,amsmath,fancyhdr,wrapfig,array,amssymb,geometry,appendix,cite}

\newcommand{\be}{\begin{equation}}
\newcommand{\ee}{\end{equation}}
\newcommand{\bea}{\begin{eqnarray}}
\newcommand{\eea}{\end{eqnarray}}
\newcommand\blfootnote[1]{%
  \begingroup
  \renewcommand\thefootnote{}\footnote{#1}%
  \addtocounter{footnote}{-1}%
  \endgroup
}

\title{Branched Hamiltonians for a class of \\Velocity Dependent Potentials}
\author[1]{Bijan Bagchi}
\author[1]{Syed M. Kamil}
\author[1]{Tarun R. Tummuru}
\author[2]{\\Iveta Semor\'{a}dov\'{a}}
\author[2]{Miloslav Znojil}
\affil[1]{Department of Physics, Shiv Nadar University, Dadri, UP 201314, India}
\affil[2]{Nuclear Physics Institute, ASCR, 250 68 $\check{R}$e$\check{z}$, Czech Republic} 
\date{\vspace{-6ex}}

\begin{document}
	\maketitle	
	
\begin{abstract}
Hamiltonians that are multivalued functions of momenta are of topical interest since they correspond to the Lagrangians containing higher-degree time derivatives. Incidentally, such classes of branched Hamiltonians lead to certain not too well understood ambiguities in the procedure of quantization. Within this framework, we pick up a model that samples the latter ambiguities and simultaneously turns out to be amenable to a transparent analytic and perturbative treatment. 
\blfootnote{
    E-mails: {bbagchi123@gmail.com, kamil.syed@snu.edu.in, tt295@snu.edu.in,\\ \hspace*{1.34cm}semoradova@ujf.cas.cz and znojil@ujf.cas.cz}}
\end{abstract}

\section{Introduction}
Models of classical systems with branched structures \cite{HTZ}, in either coordinate ($x$) space or in its momentum ($p$) counterpart, have of late been a subject of active theoretical inquiry \cite{SW,CZ,BZ,ZYX,CH,AB,RMDS,RMC}. The key idea is that classical Lagrangians possessing time derivatives in excess of quadratic powers inevitably lead to $p$ becoming a multi-valued function of velocity ($v$), thereby yielding a multi-valued class of Hamiltonian systems. 

Branched Hamiltonians in the classical context and their quantized forms have been recently discussed by Shapere and Wilczek \cite{SW}. Following it, Curtright and Zachos \cite{CZ} analyzed certain representative models for a classical Lagrangian described by a pair of convex, smoothly tied functions of $v$. Proceeding to the quantum domain shows that the double-valued Hamiltonian obtained has an inherent feature of being expressible in a supersymmetric form in the $p$ space. Subsequently, a class of nonlinear systems whose Hamiltonians exhibit branching was explored by Bagchi et al \cite{BZ} who also considered the possibility of quantization for some specific cases of the underlying coupling parameter. 

In this paper, we present a class of velocity-dependent Lagrangians which define a canonical momentum that yields exactly a pair of velocity variables. As a consequence, the corresponding Hamiltonians develop a branching character. An interesting aspect of our scheme is that it is well-suited for a perturbative treatment.

\section{Branched Hamiltonians: A brief review}
Let us briefly review the example of a branched system that was put forward in \cite{CZ}. It has been noted that a typical model of branched Hamiltonians results from a non-conventional form of the Lagrangian which is given by
\be \label{eq:1.1}  
    L = C (v - 1)^{\frac{2k-1}{2k+1}}- V(x) \text{~ where ~} C = \frac{2k+1}{2k-1} \left(\frac{1}{4}\right)^{\frac{2}{2k+1}}.
\ee
Notice that the traditional kinetic energy term features a replacement of the quadratic form by a fractional function of velocity $v$, while $V(x)$ stands for a convenient local interaction potential. The fractional powers of the difference $(v-1)$ were invoked to make plausible connections to known phenomenology such as the supersymmetric pairing. 

In the expression of canonical momentum derived from (1), the $(2k+1)-$st root of $(v-1)$ was required to be real and positive or negative for $v>1$ or $v<1$, respectively. The quantity $v$, correspondingly, turns out to be a double-valued function of $p$. Taking the Legendre transform with respect to these branches of $v$ gives rise to a pair of Hamiltonians
\be \label{eq:1.2}
    H_{\pm} = p \pm \frac{1}{4k-2} \left(\frac{1}{\sqrt{p}}\right)^{2k-1} + V(x)\,.
\ee
Note that the $k=1$ case speaks of the canonical supersymmetric structure \cite{W} for the difference $H_{\pm}-V(x)$, namely $p \pm \frac{1}{2\sqrt{p}}$, but in the momentum space if viewed as a quantum mechanical system. The spectral and boundary condition linkages of these Hamiltonians are not difficult to set up.

\section{A velocity dependent potential}
Against the above background, we consider setting up of an extended Lagrangian model having a velocity dependent potential $U(x,v)$ :
\be \label{eq:3.1}
	L(x,v) = C(v-1)^{\frac{2k-1}{2k+1}} - U(x,v) 
\ee
where $C$ is defined as in Eq.~(1). With $f(v)$ and $V(x)$ being certain functions of $v$ and $x$ respectively, we assume the potential to be given in a separable form $U(x,v) = f(v) + V(x)$. 

Using the standard definition of the canonical momentum, we find that it is given by 
\be
    p = \left(\frac{1}{4}\right)^{\frac{2}{2k+1}}(v-1)^{-\frac{2}{2k+1}} - f'(v)\,.
\ee
This relation, however, is too complicated to put down the multi-valued nature of velocity in a tractable closed form. 

If we try to determine the Hamiltonian branches corresponding to this Lagrangian (3), the $H_{\pm}$ emerge in a mixed form involving momentum $p$, the function $f(v)$ and its derivative.
\be
    H_{\pm} = p \pm \frac{1}{4}\left[p+f'(v)\right]^{-\frac{2k-1}{2}} \left(\frac{2k+1}{2k-1} - p\left[p+f'(v)\right]^{-1}\right) + U(x,v)\,.
\ee
Since the Hamiltonian has to be a function of coordinates and corresponding canonical momenta, the expression as derived above is of little use.

We note that the case $k=1$ is particularly interesting to understand the spectral properties of $L(x,v)$. Explicitly, the Lagrangian assumes a simple but general form  
\be \label{eq:3.2}
	L = 3\left(\frac{1}{4}\right)^{\frac{2}{3}} (v-1)^{\frac{1}{3}} - f(v) - V(x)\,.
\ee	
A sample choice for $f(v)$ could be 
\be \label{eq:3.3}
f(v) = \lambda v + 3\delta (v-1)^{\frac{1}{3}}
\ee
with $\lambda \left(\geq 0 \right)$ and $\delta \left(< 4^{-\frac{2}{3}}\right)$ being suitable real constants. Observe that the presence of $\delta$ rescales the kinetic energy coefficient which now enjoys a parametric representation. 

Further, this construction of $f(v)$ facilitates determination of the canonical momentum $p$ in a closed form, as given by
\be \label{eq:3.4}
	p = \mu (v-1)^{-\frac{2}{3}} - \lambda
\ee
where $\mu = 4^{-\frac{2}{3}} - \delta > 0$. On inversion, we find a pair of relations for the velocity that depend on $p$ :
\be \label{eq:3.5}
	v_{\pm}(p) = 1 \mp \mu^{\frac{3}{2}} (p + \lambda )^{-\frac{3}{2}}\,.
\ee
As a consequence, we run into two branches of the Hamiltonian which we write down as
\be	\label{eq:3.6}
	H_{\pm} - V(x) = (p+\lambda) \pm \frac{2\gamma}{\sqrt{p+\lambda}}
\ee
For the ease of notation, note that $\mu^{3/2}$ has been replaced with $\gamma$.

\begin{figure}[!ht]
    \centering
    \includegraphics[height=6cm]{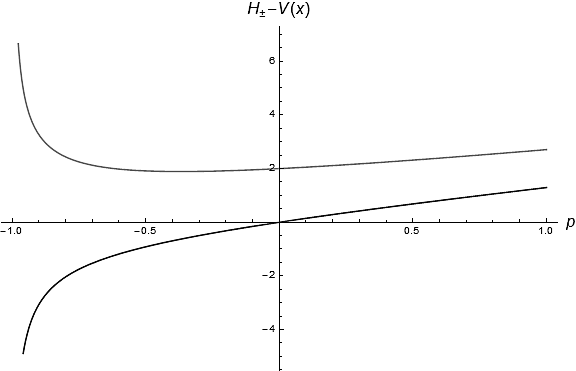}
    \caption{When $\lambda = 1$ and $\gamma=\frac{1}{2}$, $H_{\pm} - V(x)$ branches correspond to the upper and lower curves respectively.}
    \label{fig1:Ep}
\end{figure}
In the special case where $\lambda = 0$ and $\gamma = \frac{1}{4}$, we recover the Hamiltonian derived in \cite{CZ}. However, the presence of the parameter $\gamma$ in $(10)$ is nontrivial as our following treatment of perturbative analysis will show. In Figure 1, we have given a graphical illustration (for $\lambda = 1$ and $\gamma = \frac{1}{2}$) of the behavior of the two branches of the Hamiltonian against some typical values of the momentum variable. As in the $\lambda = 0$ case of \cite{CZ}, here too, for a fixed $\gamma$, we encounter a cusp asymptotically with regard to $p$.

\section{Lowest excitations and the Fourier transform}

After one decides to consider just small excitations of our quantum system over a local or global minimum ($x_0$) of a generic analytic potential $V(x)$, one may put the origin of the coordinate axis to this minimum, $x \to y=x-x_0$, and write down the Taylor series
\be
		V(x) = V(x_0) + (x-x_0) V'(x_0) + \frac{1}{2} (x-x_0)^2 V''(x_0)+\ldots
		\label{eq:4.2}
\ee
Recall that $V'(x_0) = 0$ and the zero of the energy scale can be shifted in such a manner that $V(x_0) = 0$. Finally, the series is truncated after the first non-trivial term yielding, in \textit{ad hoc} units,
\be
	V(x_0+y) = y^2.
	\label{eq:4.3}
\ee

After a Fourier transform to the momentum space, we get a transformed quantum form of the Hamiltonian guided by the second-order differential operator,
\be	
	H = -\frac{d^2}{dp^2} + W(p)
	\label{eq:4.4}
\ee
containing a one-parametric family of pseudo-potentials
 \be
    W(p) = p + \frac{2\gamma}{\sqrt{p}}\,.
    \label{eq:4.5}
 \ee
Here, the original subscript $_\pm$ entering Eq.~(10) may be perceived as equivalent to an optional switch between positive coupling-type parameter $\gamma>0$ and its negative alternative $\gamma<0$. Besides such a freedom of the sign of the dynamical characteristic, the consequent quantum-theory interpretation of the model requires also a few nontrivial mathematical addenda. The form Eq.~(14) matches with Eq.~(10) for $\lambda = 0$ which will now be our point of inquiry.

First of all, the most natural tentative candidate
 \be
    H \phi_n(p) = E_n \phi_n(p)\,,\ \ \ \ \ p \in (-\infty,\infty)
    \label{eq:4.6b}
 \ee
for the quantum Schr\"{o}dinger equation living on the whole real line of momenta (i.e., with $\phi_n(p) \in L^2 (\mathbb{R})$) is characterized by the asymptotically linear {\em decrease} of the pseudo-potential (14) along the left half-line. Hence, the negative half-axis of momenta $p$ must be excluded, {\it a priori}, as unphysical. In other words, the acceptable wave functions $\phi_n(p)$ should vanish, identically, whenever $p \in (-\infty,0)$. The consistent quantization of our model must be based on the modified, half-line version of Eq.~(15), viz., on Schr\"{o}dinger equation
\be
    H \phi_n(p) = E_n \phi_n(p)\,, \hspace{0.5cm} p \in (0,\infty)
    \label{eq:4.6}
\ee
such that (cf. also \cite{SW} and \cite{CZ})
\be
 \phi_n(p) \in L^2 (\mathbb{R}^+)\,.
    \label{eq:4.6bc}
\ee
Still, the discussion is not yet complete. Due care must be also paid to the fact that the inverse-square-root singularity of $W(p)$ in the origin is ``weak'' (see, e.g., Ref.~\cite{cond} for a detailed explanation of the rigorous, ``extension theory'' mathematical contents of this concept). In the language of physics, such a comment means that the information about possible bound states and physics represented by Eq.~(16) with constraint
(17) is incomplete.

In the rest of this paper (i.e., in sections \ref{reperros} and \ref{perros}) we shall, therefore, describe the two alternative versions of the completion of the missing, phenomenology-representing information.

\section{Eligible ``missing'' boundary conditions at small $\gamma$
and $p=0$}\label{reperros}

As we emphasized above, the existence of the usual discrete spectrum of bound states can only be guaranteed via an {\em additional} physical boundary condition at $p=0$. Although from the point of view of pure mathematics, the choice of such a condition is flexible and more or less arbitrary, the necessary suppression of this unwanted freedom can rely upon several forms of the physics-based intuition.

Let us split the problem into two subcategories. In a simpler scenario we shall assume that the central core is repulsive and strong (i.e., that our parameter is positive and large, $\gamma \gg 1$). This possibility will be discussed in the next section \ref{perros}. For the present, let us admit that the (real) value of $\gamma$ is arbitrary and that the regular nature of our ordinary differential Schr\"{o}dinger equation near $p=0$ implies that the integrability condition (17) itself still does not impose any constraint upon the energy $E$ \cite{cond}. A fully explicit and constructive demonstration of such an observation may be based on the routine reduction of (16) to its simplified, leading-order form
 \be
 -\sqrt{p}\,\frac{d^2}{dp^2}\psi(p)+ 2\gamma\,\psi(p)=0\,.
 \ee
Being valid at the very small (though still positive) values of $p \ll 1$ this equation is exactly solvable in terms of Bessel functions \cite{nist}. Thus, one may choose either $\gamma>0$ or $\gamma<0$.

After some algebra, we obtain the respective two-parametric families of the general  solutions which depend on two parameters $C_{1,2}$ or $D_{1,2}$ and which remain energy-dependent. At small $p$ they behave, respectively, as follows,
\be
 \psi \left( p \right) =   C_1\,\sqrt{p}\,{\it I}_{2/3} \left(\frac{4\sqrt{2}}{3} \sqrt{\gamma}\,{p}^{\frac{3}{4}} \right) + C_2\,\sqrt {p}\,{\it K}_{2/3} \left(\frac{4\sqrt{2}}{3} \sqrt{\gamma}\,{p}^{\frac{3}{4}} \right)
\ee
and
\be
 \psi \left( p \right) =
 D_1\,\sqrt {p}\,{\it J}_{2/3} \left(\frac{4\sqrt{2}}{3} \sqrt{-\gamma}\,{p}^{\frac{3}{4}} \right) + D_2\,\sqrt {p}\,{\it Y}_{2/3} \left( \frac{4\sqrt{2}}{3} \sqrt{-\gamma}\,{p}^{\frac{3}{4}} \right).
\ee
On this purely analytic background, one of the most natural resolutions of the paradox of the ambiguity of the physical boundary conditions at $p=0$  may be based on the brute-force choice of the parameters $C_{1,2}$ or $D_{1,2}$ in these formulae.

Finally, let us emphasize that intuitively by far the most plausible requirement of the absence of the jump in the wave functions at $p=0$, i.e., the Dirichlet boundary condition 
\be
 \lim_{p\to 0} \psi \left( p \right) =0
 \label{Diri}
\ee
would remove the latter ambiguity of quantization in the most natural manner. The resulting pair of the requirements
 \be
 C_2=0,\hspace{0.3cm} D_2=0
 \ee
may be then recommended as easily derived from the well known approximate formulae for the Bessel functions near the origin
\cite{nist}.

\section{Perturbation-theory analysis at large $\gamma \gg 1$}\label{perros}
In a purely formal spirit, one could complement the above recommended Dirichlet boundary condition (21) by its Neumann vanishing-derivative analogue
 \be
 \lim_{p\to 0} \psi' \left( p \right) =0
 \label{Neum}
 \ee
or, more generally, by a suitable Robin boundary condition. In this context it is worth adding that with a systematic strengthening of the repulsive version of the barrier (i.e., with the growth of the positive coupling constant $\gamma $) the specification of the additional boundary conditions at $p=0$ becomes less and less relevant because the two alternative energy levels will degenerate in the limit $\gamma \to \infty$.

The most immediate explanation of this phenomenon may be provided by perturbation theory. In the dynamical regime, when the parameter is large, a perturbative approach seems to be particularly well suited. With $\gamma \gg 1$, we look at the absolute minimum of the potential $W(p)$ which occurs at $p_0$, say. This value is, incidentally, unique
\be
	p_0 = \gamma^{\frac{2}{3}} \gg 1
\ee
With the construction of a Taylor series in its vicinity, 
\be
	W(p) = W(p_0) + (p-p_0) W'(x_0) + \frac{1}{2} (p-p_0)^2 W''(p_0)+\ldots
\ee
we observe that the first term, which is given by
\be
	W(p_o) = 3 \gamma^{\frac{2}{3}}
\ee
in very large in this scenario. In contrast, all of the further Taylor coefficients remain very small and asymptotically negligible, 
\be
	W''(p_o) = \frac{3}{2} \gamma^{-\frac{2}{3}}, \hspace{0.3cm} W'''(p_o) = -\frac{15}{4} \gamma^{-\frac{4}{3}} \hspace{0.3cm} \ldots
\ee
Clearly then, with $\gamma \gg 1$, $H$ can be expressed as 
\be	
	H = -\frac{d^2}{dp^2} + 3 \gamma^{\frac{2}{3}} + \frac{3}{4} \gamma^{-\frac{2}{3}} (p-p_0)^2 + \mathcal{O}(\gamma^{-\frac{4}{3}} (p-p_0)^3)\,.
\ee
After one re-scales the axis $p = \rho q$, Eq.~(16) acquires the modified form 
\be
	\tilde{H} \tilde{\phi}_n(q) = E_n \rho^2 \tilde{\phi}_n(q)
\ee
where,
\be
	\tilde{H} = -\frac{d^2}{dq^2} + 3\rho^2 \gamma^{\frac{2}{3}} + \frac{3}{4}\rho^4 \gamma^{-\frac{2}{3}} (q-q_0)^2 + \mathcal{O}(\rho^5\gamma^{-\frac{4}{3}} (q-q_0)^3)\,.
\ee
One may now set 
\be	
	\rho = \left(\frac{4}{3}\right)^{\frac{1}{4}} \gamma^{\frac{1}{6}}
\ee
yielding the very weakly perturbed harmonic-oscillator Hamiltonian
\be
	\tilde{H} = -\frac{d^2}{dq^2} + (q-q_0)^2 + \gamma \sqrt{12} + \mathcal{O}(\rho^5\gamma^{-\frac{4}{3}} (q-q_0)^3)\,.
\ee
In full analogy to many models with similar structure (cf., the study \cite{MZ} containing further references), the exact solvability of the model in the leading-order harmonic oscillator approximation proves sufficient because in the domain of large $\gamma \gg 1$ the contribution of the anharmonic corrections becomes negligible.

\section{Summary}

To summarize, we looked at the particular example of a non-conventional Lagrangian with a velocity-dependent potential that leads to a double-valued structure of the associated Hamiltonian for some specific choice of the underlying coupling parameter. We showed that our scheme allows for a perturbative analysis by constructing a Taylor series near the vicinity of the absolute minimum of the potential.

\section{Acknowledgments}
For one of us (BB), it is a pleasure to thank Prof. Sara Cruz y Cruz and Prof. Oscar Rosas-Ortiz for warm hospitality during Quantum Fest 2016 held at UPIITA-IPN, Mexico City. MZ acknowledges the short-stay hospitality by Shiv Nadar University and the support by the GA\v{C}R Grant Nr. 16-22945S. We also all thank Prof. Bhabani Prasad Mandal for fruitful discussions.

\end{document}